\documentclass[preprint,showpacs,preprintnumbers,amsmath,amssymb]{revtex4}

\usepackage{graphicx}
\usepackage{dcolumn}
\usepackage{bm}

\begin{document}

\preprint{submitted to Phys. Rev. Lett.}

\title{Tamm oscillations in semi-infinite nonlinear waveguide arrays}

\author{Milutin Stepi\'c}
 \altaffiliation[Also at ]{Vin\v ca
Institute of Nuclear Sciences, P.O.B. 522, 11001 Belgrade,
Serbia.} \email{milutin.stepic@tu-clausthal.de}

\author{Eugene Smirnov}
\author{Christian E. R\"uter}
\author{Detlef Kip}%
 \affiliation{Institute of Physics and Physical Technologies, Clausthal University of Technology, 38678 Clausthal-Zellerfeld, Germany
}%

\author{Aleksandra Maluckov}
\affiliation{Faculty of Sciences and Mathematics, University of
Ni\v s, P.O. Box 224, 18000 Ni\v s, Serbia}

\author{Ljup\v co Had\v zievski}
\affiliation{Vin\v ca Institute of Nuclear Sciences, P.O.B. 522,
11001 Belgrade, Serbia}

\date{\today}

\begin{abstract}
We demonstrate numerically the existence of nonlinear Tamm
oscillations at the interface between a substrate and
one-dimensional waveguide array with both cubic and saturable,
self-focusing and self-defocusing nonlinearity. Light is trapped
in the vicinity of the boundary of the array due to the interplay
between the repulsive edge potential and Bragg reflection inside
the array. In the special case when this potential is linear these
oscillations reduce themselves to surface Bloch oscillations.
\end{abstract}


\maketitle

The problem of surface waves which could exist at the interface
between two different media has been studied for several decades.
These waves are, for example, investigated at a metal-vacuum
interface \cite{1}, air-water interface \cite{2}, at the interface
between two dielectrics \cite{3}, and in finite anti-dot lattices
\cite{4}. Tamm was the first to consider neglected edge effects in
a semi-infinite Kronig-Penney model \cite{5}. He has discovered
that strongly localized surface states could appear providing that
the surface potential perturbation is strong enough. However, due
to various experimental difficulties, his discovery was awaiting
more than half a century on the experimental confirmation in
semiconductor superlattices \cite{6}. From the practical point of
view, surface waves represent convenient tools to investigate the
properties of material interfaces. Furthermore, they could be
implemented in various devices such as bio-sensors \cite{7} and
polariton lasers \cite{8}.

Recently it has been suggested that these surface states may also
exist at the interface between a homogenous medium (substrate) and
a nonlinear waveguide array (NWA) \cite{9}. NWA represent arrays
of evanescently coupled optical waveguides. Up to date, NWA have
been successively fabricated in materials exhibiting cubic
\cite{10,11,12}, quadratic \cite{13}, saturable \cite{14,15}, and
nonlocal nonlinearity \cite{16}. Discrete solitons
\cite{10,13,14,17}, breathers \cite{18}, diffraction management
\cite{11}, modulational instability \cite{12,15,16}, and Bloch
oscillations \cite{19} are just a few examples of phenomena which
have been observed in such systems.

Very recently, the first experimental observation of discrete
surface solitons in AlGaAs NWA exhibiting a cubic self-focusing
nonlinearity \cite{20} has triggered further investigations of
surface waves at the interface between a NWA and a substrate. The
existence of surface gap solitons in the lattice with cubic
self-defocusing nonlinearity has been reported in Ref.~\cite{21}.
A crossover from nonlinear surface states to discrete solitons was
studied, too \cite{22}. In these two papers it has been revealed
that the vicinity of the edge enables a stable propagation of
various localized modes, such as flat-top modes and inter-site
ones (mode B) \cite{23}. Very recently, strongly localized surface
waves have been independently observed in NWA exhibiting saturable
\cite{24,25} and quadratic nonlinearity \cite{26}, respectively.
The fact that a surface can support the existence of localized
states has been, for example, further exploited in vector
\cite{27}, nonlocal \cite{28} as well as binary NWA \cite{29}, and
in a soliton array case \cite{30}.

In this Letter we reveal the existence of Tamm oscillations at the
edge of a semi-infinite NWA. These oscillations are the result of
an interplay between a repulsive potential which originates from
the boundary of the array and the array's periodicity. We
calculate this potential for a few different media and reveal that
Tamm oscillations are more likely to occur in systems with
stronger coupling. In a special case when the repulsive potential
is a linear function of the distance from the edge of the NWA Tamm
oscillations reduce to the well-known Bloch oscillations
\cite{31,32,33}, which have considerable potential in all-optical
switching, amplification, and steering \cite{19,32}.

The propagation of light in periodically stratified structures
such as NWA can be described analytically either by the
Floquet-Bloch model \cite{15,34} or by a coupled-wave theory
\cite{9,10,13,17}. The model equation within the first approach
reads:
\begin{equation}
i\frac{\partial E}{\partial y}+ \frac{1}{2k}\frac{\partial^2
E}{\partial z^2}+k\frac{n(z)+\Delta n_{nl}}{n_s}E=0~.
\end{equation}
The propagation coordinate is along the $y$-axis, the amplitude of
the electrical field is denoted by $E$, while $k=2\pi n_s/\lambda$
represents the wave number. Here, $\lambda$ is the wavelength of
the used light in vacuum, and $n_s$ is the extraordinary
refractive index of the substrate. The periodical modulation of
the refractive index which defines the nonlinear WA is denoted by
$n(z)$ while $\Delta n_{nl}$ is the nonlinear refractive index
change ($\Delta n_{nl}<<n_s$). In media with cubic nonlinearity
$\Delta n_{nl}=\Delta n_0 I$ while in media with saturable
nonlinearity we have $\Delta n_{nl}=\Delta n_0 I/(I+I_d)$, where
$I$ is the peak light intensity and $I_d$ is the so-called dark
irradiance \cite{14}.

On the other hand, light propagation through NWA may be described,
within the tight-binding approximation, by the following set of
conveniently normalized, linearly coupled, nonlinear, ordinary
differential equations:
\begin{equation}
i\frac{d U_n}{d \xi}+C(U_{n+1}+U_{n-1}-2U_{n})- g(|U_n|^2)U_n=0,%
\end{equation}
where $g(|U_n|^2)=\beta|U_n|^2$ in cubic ($c$) media and
$g(|U_n|^2)=\beta|U_n|^2/(1+|U_n|^2)$ in saturable ($s$) media
\cite{35}, while the nonlinearity coefficient $\beta<0$ for the
self-focusing ($f$) and $\beta>0$ for the self-defocusing ($d$)
case, respectively. Here $C$ is the coupling constant, $\xi$ is
the propagation coordinate and $U_n$ is the normalized electric
field envelope in the $n$-th waveguide \cite{36}. Integral of
motions are power and Hamiltonian:
\begin{eqnarray}
P&=&\sum_{n}|U_n|^2, \nonumber\\
H_{s}&=&\sum_{n}{\{C|U_{n-1}-U_n|^{2} - \beta
[\ln{(1+|U_n|^2)}-|U_n|^2]\}}, \nonumber\\
H_{c}&=&\sum_{n}[C|U_{n-1}-U_n|^{2}+ \frac{\beta |U_n|^4}{2}].
\end{eqnarray}
Assuming stationary solutions of staggered form $U_n=F_n
\exp{[i(-\nu \xi +n\pi)]}$ ($\nu$ represents soliton frequency)
for defocusing cases, and of unstaggered form $U_n=F_n \exp{(-i\nu
\xi)}$ for focusing cases, together with the assumption $|F_0|\gg
|F_{\pm 1}|\gg |F_{\pm 2}|$ for on-site (A) mode and $|F_{\pm
1}|\gg |F_{\pm 2}|\gg |F_{\pm 3}|$ for inter-site (B) mode, one
may find the following expressions for the maximal amplitude in
the array:
\begin{eqnarray}
F_{0s(f,d)}&=&\sqrt{\frac{\nu - 2C}{\beta-\nu + 2C}},\;
F_{0c(f,d)}=\sqrt{\frac{\nu -2C}{\beta}},\\
F_{1s(f)}&=&\sqrt{\frac{\nu - C}{\beta-\nu + C}},\
F_{1s(d)}=\sqrt{\frac{\nu - 3C}{\beta-\nu + 3C}},\\
F_{1c(f)}&=&\sqrt{\frac{\nu -3C}{\beta}},\,
F_{1c(d)}=\sqrt{\frac{\nu - C}{\beta}}.
\end{eqnarray}
In addition for A mode $F_{n(>0)}=\alpha^n F_{0}$ and for B mode
$F_{n(>1)}=\alpha^{n-1} F_{1}$, in both saturable and cubic media.
Here $\alpha=\pm C/(\nu-2C)$ for $(d)$ and $(f)$ case,
respectively. Examples of the oscillatory behavior of both on-site
mode (mode A) and mode B are depicted in Fig.~1. Here the energy
is too low to overcome the repulsive potential from the edge of
the array so both modes start to move away from the interface
until they are back-reflected because of the Bragg condition.  As
this traversing is usually accompanied with radiation reflected
modes loose power and eventually do not reach back to the first
channel of the array. As a result, consequent oscillations have
longer and longer periods and modes gradually run away from the
edge. The trapping of localized modes at one channel after one or
more oscillations is possible as well. Here is interesting to
mention that surface waves could exist due to the balance between
self-bending and deflection from the edge of a bulk
photorefractive crystal, too \cite{37}. Equivalently, the behavior
of moving localized modes can be interpreted by the interplay
between a repulsive force (due to the boundary effects) and the
effective Peierls-Nabarro potential (due to system discreteness)
\cite{35}.

\begin{figure}[tb]
\includegraphics[width=8.0cm]{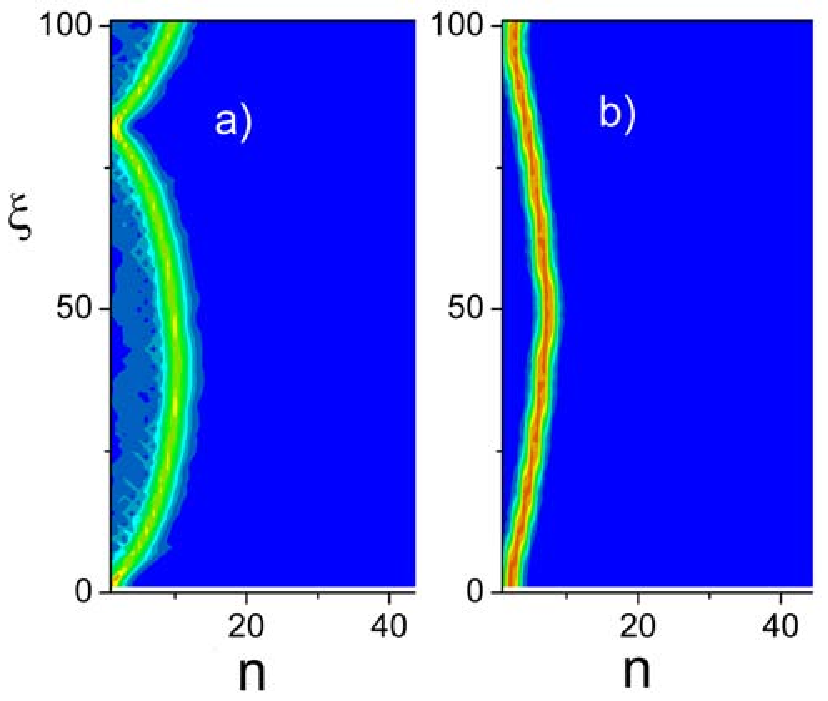}
\caption{\label{Fig1} Tamm oscillations of a) mode A with $C=0.5$,
$|\beta|=3.34$, and soliton frequency $\nu=2.3$ launched into the
first channel of the array, and b) mode B with $C=0.5$,
$|\beta|=3.34$, and soliton frequency $\nu=4.83$ launched into the
second and third channel of the array.}
\end{figure}

We calculate the effective repulsive potential as the difference
between the truncated potential and a reference potential with
periodic boundary condition, and present results for symmetric
localized structures centered on-site in an array that consists of
13 elements. As the peak of this localized structure (with the
highest amplitude $F_0$) approaches the edge more and more terms
of the corresponding Hamiltonian have to be truncated. For
example, if the peak of mode A resides in the 6th channel one can
obtain the following expressions for the repulsive potential:
\begin{eqnarray}
V_{rep.c}^{(1)}&=&
C|F_0|^{2}\alpha^{10}-C|F_0|^{2}(1+\alpha)^2\alpha^{10}-0.5\beta|F_0|^{4}
\alpha^{24}, \nonumber \\
V_{rep.s}^{(1)}&=&C|F_0|^{2}\alpha^{10}-C|F_0|^{2}(1+\alpha)^2\alpha^{10}\nonumber
\\& & -\beta |F_0|^{2} \alpha^{12}+\beta\,
ln(1+\alpha^{12}|F_0|^{2}),
\end{eqnarray}
in both saturable and cubic media in defocusing and focusing
cases. After a straightforward calculation it is possible to get
explicit expressions for $V_{rep}^{(2)}, V_{rep}^{(3)},$ etc.\
where the number in the superscript denotes the number of the
truncated channels. For example, for the Tamm state in the
self-focusing cubic case we have found:
\begin{eqnarray}
V_{rep.cf}^{(6)}&=&C|F_0|^{2}\nonumber
\\&-C&|F_0|^{2}(1+\alpha)^2(\alpha^{10}+\alpha^{8}+\alpha^{6}+\alpha^{4}+\alpha^{2}+1)\nonumber
\\&-0.5&\beta|F_0|^{4}
(\alpha^{24}+\alpha^{20}+\alpha^{16}+\alpha^{12}+\alpha^{8}+\alpha^{4}).
\end{eqnarray}
Note that here $\beta<0$ and $\alpha=-C/(\nu-2C)$. The dependence
of the repulsive potential on the distance from the edge of the
array is presented in Fig.~2. Stronger coupling (i.e.\ shorter
coupling length) results in stronger repulsion (see Fig.~2a) while
stronger nonlinearity decreases the repulsive potential (Fig.~2b).
Beams which are strongly pushed off from the edge will experience
Bragg reflection inside the array earlier than weakly rejected
beams. Thus, the same input beam will have shorter spatial periods
of Tamm oscillations in arrays with stronger coupling and weaker
nonlinearity.

\begin{figure}
\includegraphics[width=8.0cm]{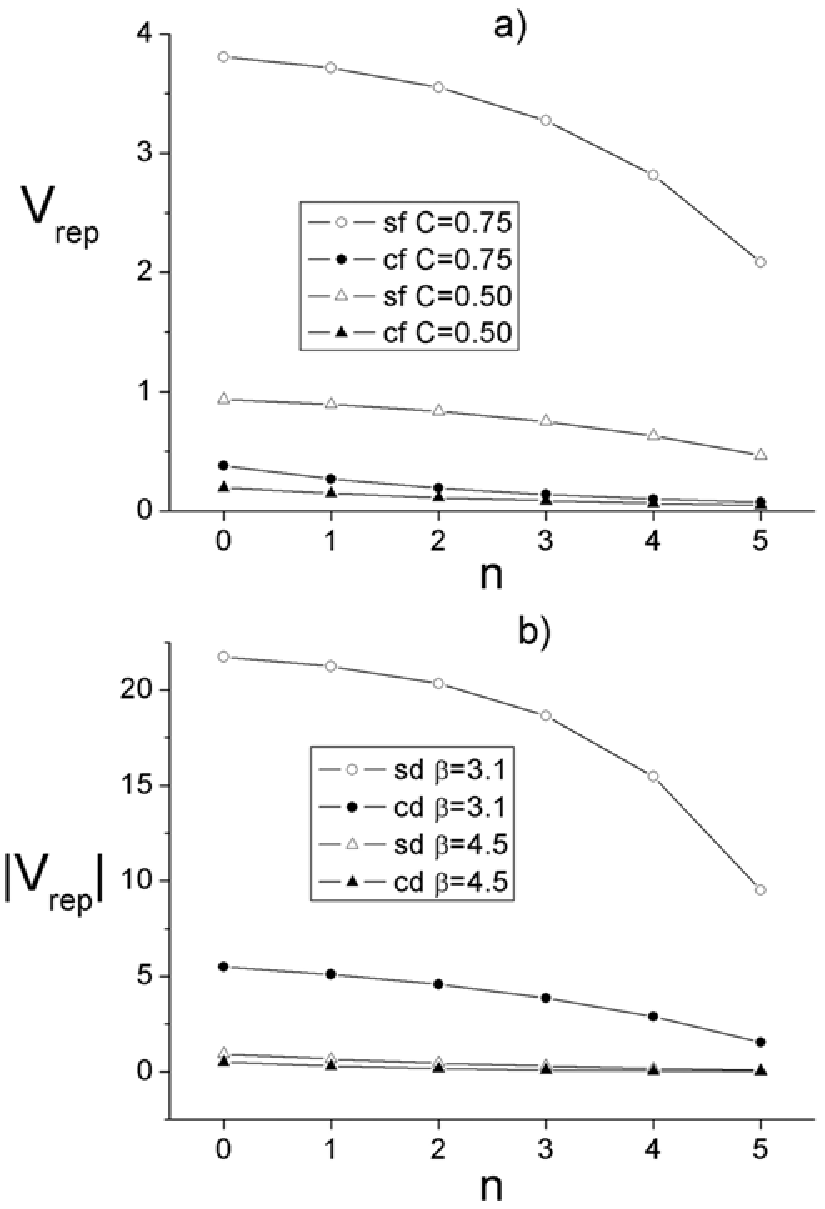}
\caption{\label{Fig2} Dependence of the effective repulsive
potential on the distance from the interface between a substrate
and the array. a) Self-focusing case: $|\beta|=3.1$,
$|F_0|^2=0.24$ and different values of $C$. b) Self-defocusing
case: $C=0.5$, $|F_0|^2=0.172$, and different values of $\beta$.}
\end{figure}

The value of the corresponding repulsive potentials for both cubic
and saturable, self-focusing and self-defocusing cases also
depends on the soliton frequency $\nu$. For a fixed value of this
frequency and fixed values of parameters $C$ and $\beta$, the
following relation is usually fulfilled:
$V_{rep}^{sd}>V_{rep}^{cd}>V_{rep}^{sf}>V_{rep}^{cf}$ but interior
elements may permute their position as well. This relation may
explain why this effect has not been observed in recent studies in
AlGaAs waveguide arrays exhibiting a cubic self-focusing
nonlinearity \cite{9,20}. Interestingly, in the case when
$V_{rep}$ is a linear function from the interface distance the
input beam will experience Bloch oscillations \cite{19}, which
have a considerable potential for application in all-optical
devices as reported, for example, in Ref.~32. Restricting
ourselves here only to NWA, this phenomenon was studied in curved
NWA \cite{33}, and in NWA with a linearly growing effective
refractive index \cite{19}. Please note that, from the surface's
point of view, Tamm oscillations do not belong to nonlinear
effects (i.e.\ they exist for light intensities lower than the
threshold for the onset of highly nonlinear Tamm states
\cite{9,20,21,22,24,25,26}).

In order to check our findings we performed additional simulations
based on a numerical solution of Eq.~(1). We, arbitrarily, have
used the parameters which are achievable in lithium niobate
waveguide arrays exhibiting a self-defocusing saturable
nonlinearity ($\Delta n_0=3.7\times 10^{-4}$, $\Delta n_{nl}\le
0.001$) and green light with $\lambda=532\,$nm. For such samples
the periodically modulated refractive index can be well
approximated by a $\cos^2$ function \cite{15}. Numerical results
which have been obtained by virtue of a beam propagation method
are given in Fig.~3. Note that in reality this effect occurs in a
highly elongated space (20\,mm$\times 0.1\,$mm in Fig.~3a).

\begin{figure}
\includegraphics[width=8.0cm]{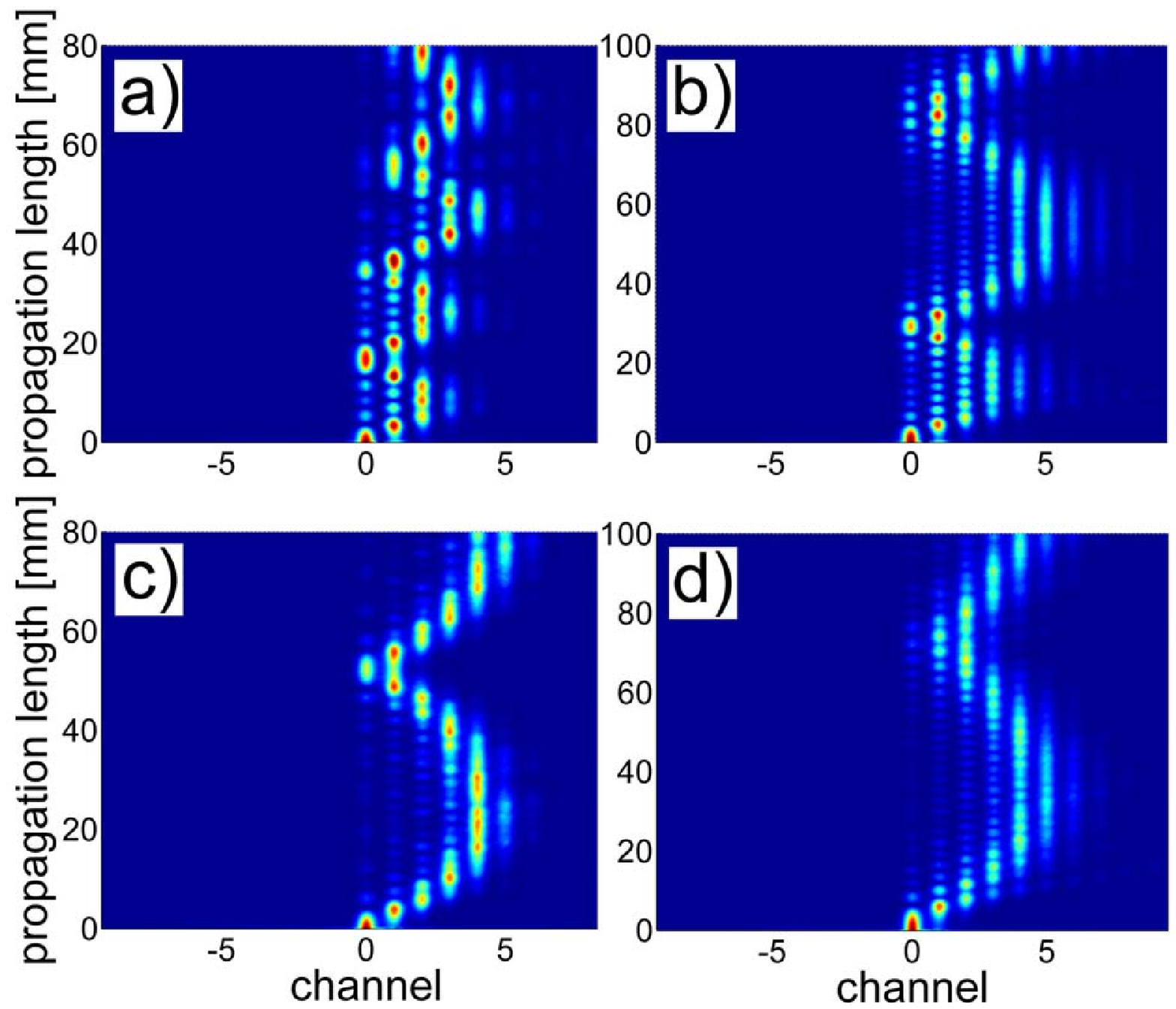}
\caption{\label{Fig3} (Color online) Tamm oscillations in a
waveguide array where channels are 4\,$\mu$m wide and separated by
4.4\,$\mu$m while $\Delta n_0=3.7\times 10^{-4}$. The input
pattern with amplitude ratios of $1:0.5:0.1$ has been launched
into the first channel. a) Self-defocusing saturable case $\Delta
n_{nl}=6.2\times 10^{-4}$, $r=I/I_d=6$, b) self-defocusing cubic
case $\Delta n_{nl}=4.42\times 10^{-4}$, c) self-focusing
saturable case $\Delta n_{nl}=3.34\times 10^{-4}$, $r=6$, and d)
self-focusing cubic case $\Delta n_{nl}=2.62\times 10^{-4}$.}
\end{figure}

As the position of the input beam shifts towards the interior of
the array, the period of Tamm oscillations increases. Fig.~4 may
be understood as a numerical proof that these oscillations are
indeed a linear surface effect. Namely, the light intensity
necessary to form Tamm oscillations (Fig.~4a) is not high enough
to form a surface soliton \cite{9,20,21,22,24,25,26} but suffices
to form a narrow breather in a channel which is only two channels
away from the substrate-array interface (Fig.~4b).

\begin{figure}
\includegraphics[width=8.0cm]{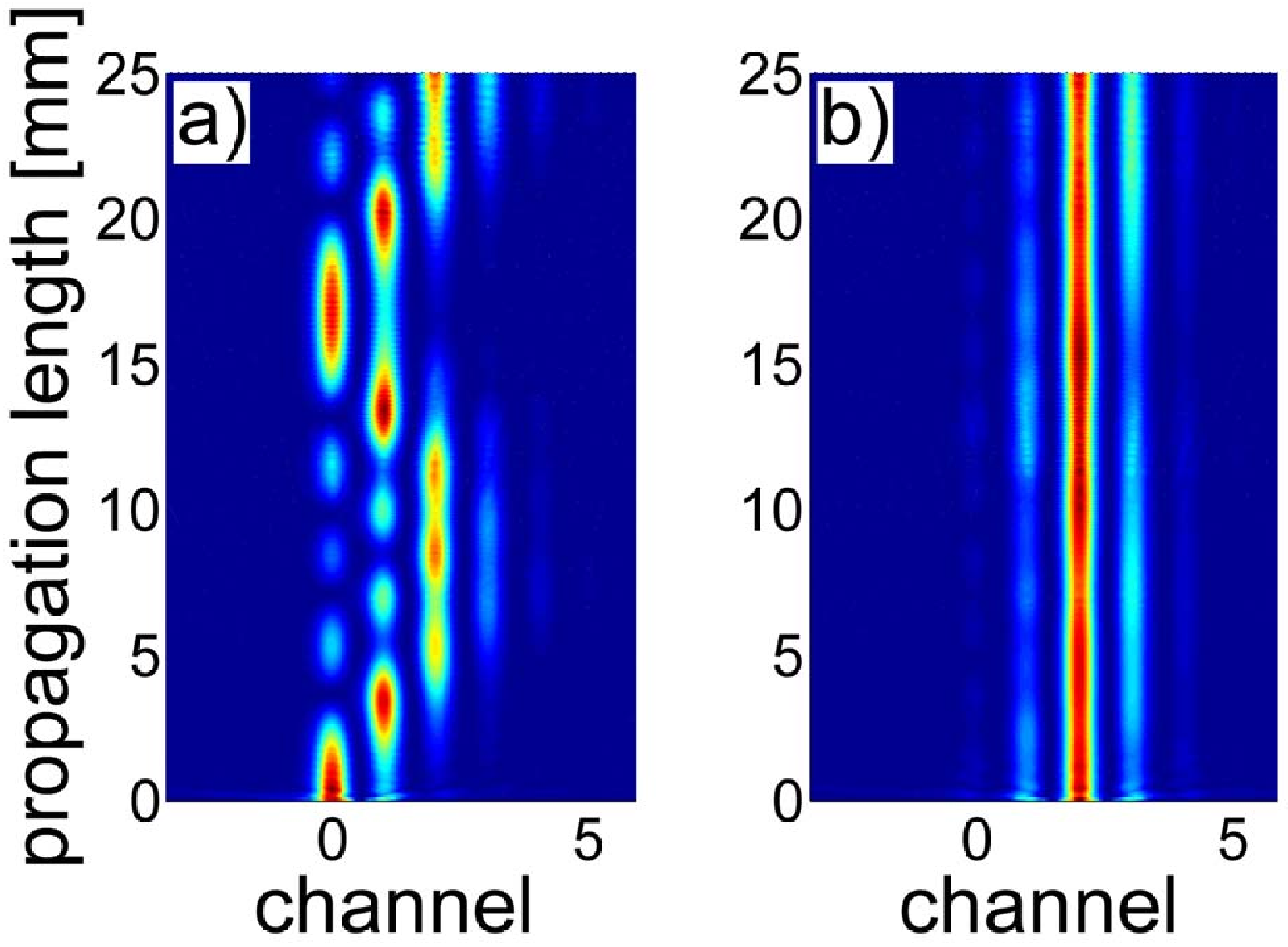}
\caption{\label{Fig4} (Color online) Light propagation in a
saturable nonlinear NWA with the same input profile and lattice
parameters as in Fig.~3. Here $\Delta n_{nl}=6.21\times 10^{-4}$
and $r=6$. The input beam is launched into a) the first channel
and b) into the third channel of a semi-infinite lattice.}
\end{figure}

Experimentally, Tamm oscillations may be observed, for example, in
one-dimensional lithium niobate waveguide arrays, where sample
lengths up to 50\,mm have been achieved \cite{24,26}. However, up
to now we failed in a direct observation of Tamm oscillations from
the top of the sample because of the rather low average level of
scattered light that is superimposed by larger scattering
amplitudes from small surface defects. An alternative approach has
been outlined in Ref.~\cite{19}, and corresponding experiments may
be performed in the near future.

In conclusion, we have demonstrated numerically the existence of
Tamm oscillations at the interface between a substrate and a
one-dimensional homogeneous nonlinear waveguide array. Light is
trapped in the vicinity of the edge of the array due to the
interplay between the edge repulsion and Bragg reflection.
Approximate analytical expressions for the repulsive truncated
potential are given for different types of nonlinear interactions
as well as for different system parameters. These oscillations
reduce to the case of Bloch oscillations when the repulsive
potential is a linearly decreasing function of the distance from
the edge of the semi-infinite nonlinear waveguide array.

\medskip
\begin{acknowledgments}
This work has been supported by the German Federal Ministry of
Education and Research (BMBF, grant DIP-E6.1), the German Research
Foundation (DFG, grant KI482/8-1)), and the Ministry of Science
and Environmental Protection of Republic Serbia, Project 14-1034.
\end{acknowledgments}

\bibliography{apssamp}

\begin{thebibliography}{}
\bibitem{1} A. Otto, Zeit. Phys. {\bf 216,} 398 (1968).
\bibitem{2} D. Middleton and R. H. Mellen, J. Acoust. Soc. Am. {\bf 90,} 741 (1991).
\bibitem{3} W. J. Tomlinson, Opt. Lett. {\bf 5,} 323 (1980).
\bibitem{4} P. H. Rivera et al., \prb {\bf 64,} 035313 (2001).
\bibitem{5} I. Tamm, Phys. Z. Sowjetunion {\bf 1,} 733 (1932).
\bibitem{6} H. Ohno et al., \prl {\bf 64,} 2555 (1990).
\bibitem{7} C. Chou et al., Opt. Exp. {\bf 14,} 4307 (2006).
\bibitem{8} A. Kavokin, I. Shelykh, and G. Malpuech, Appl. Phys. Lett. {\bf 87,} 261105 (2005).
\bibitem{9} K. G. Makris et al., Opt. Lett. {\bf 30,} 2466 (2005).
\bibitem{10} H. Eisenberg et al., \prl {\bf 81,} 3383 (1998).
\bibitem{11} H. Eisenberg et al., \prl {\bf 85,} 1863 (2000).
\bibitem{12} J. Meier et al., \prl {\bf 92,} 163902 (2004).
\bibitem{13} R. Iwanow et al., \prl {\bf 93,} 113902 (2004).
\bibitem{14} N. K. Efremidis et al., \pre {\bf 66,} 046602 (2002).
\bibitem{15} M. Stepi\'c et al., \ol {\bf 31,} 247 (2006).
\bibitem{16} M. Peccianti et al., Nature {\bf 432,} 733 (2004).
\bibitem{17} D. N. Christodoulides and R. I. Joseph, \ol {\bf 13,} 794 (1988).
\bibitem{18} S. Flach and C. R. Willis, Phys. Rept. {\bf 295,} 181 (1998).
\bibitem{19} R. Morandotti et al., \prl {\bf 83,} 4756 (1999).
\bibitem{20} S. Suntsov et al., \prl {\bf 96,} 063901 (2006).
\bibitem{21} Y. V. Kartashov, V. A. Vysloukh, and L. Torner, Phys. Rev. Lett. {\bf 96,} 073901 (2006).
\bibitem{22} M. I. Molina, R. A. Vicencio, and Yu. S. Kivshar, \ol {\bf 31,} 1693 (2006).
\bibitem{23} F. Lederer, S. Darmanyan and A. Kobyakov, Discrete solitons, in {\it Spatial solitons}, ed.\ by S. Trillo and W. Torruellas (Springer, Berlin 2001), p.~269.
\bibitem{24} E. Smirnov et al., \ol {\bf 31}, 2338 (2006).
\bibitem{25} C. Rosberg et al., arXiv:nlin.PS/0603202 (2006).
\bibitem{26} G. A. Siviloglou et al., Opt. Exp. {\bf 14,} 5508 (2006).
\bibitem{27} J. Hudock et al., Opt. Exp. {\bf 13,} 7720 (2005).
\bibitem{28} Y. V. Kartashov et al., arXiv:nlin.PS/0606160 (2006).
\bibitem{29} M. I. Molina et al., \ol {\bf 31,} 2332 (2006).
\bibitem{30} Y. V. Kartashov et al., \ol {\bf 31,} 2329 (2006).
\bibitem{31} F. Bloch, Z. Phys. {\bf 52,} 555 (1928).
\bibitem{32} U. Peschel, T. Pertsch, and F. Lederer, \ol {\bf 23,} 1701 (1998).
\bibitem{33} N. Chiodo et al., \ol {\bf 31,} 1651 (2006).
\bibitem{34} P. Yeh, A. Yariv and C.-S. Hong, J. Opt. Soc. Am. {\bf 67,} 423 (1977).
\bibitem{35} Lj. Hadzievski et al., Phys. Rev. Lett. {\bf 93,} 033901 (2004).
\bibitem{36} M. Stepi\'c et al., to be published in Opt. Commun. (2006).
\bibitem{37} M. Cronin-Golomb. \ol {\bf 20,} 2075 (1995).
\end{thebibliography}
{}
\end{document}